\documentclass[12pt,english,floatfix,superscriptaddress,aps,prd,preprint]{revtex4}
\usepackage[utf8]{inputenc}
\usepackage{amsmath}
\usepackage{amssymb}
\usepackage{amsbsy}
\usepackage{amsfonts}
\usepackage{amsopn}
\usepackage{amstext}
\usepackage{graphicx}
\usepackage{amssymb}
\usepackage{amsfonts}
\usepackage{amsmath}
\usepackage{amsmath,amsthm,amsfonts,amssymb}
\usepackage[mathcal]{eucal}
\usepackage{mathrsfs}
\usepackage{graphicx}
\usepackage[english]{babel}
\usepackage{color}
\usepackage{esint}
\usepackage[dvips]{epsfig}
\usepackage[dvips]{graphicx}
\usepackage{float}
\usepackage{units}
\usepackage{textcomp}

\usepackage{hyperref}
\usepackage{slashed}

\newcommand{\ie}{\begin{equation}}
\newcommand{\fe}{\end{equation}}
\newcommand{\se}{\begin{eqnarray}}
\newcommand{\ff}{\end{eqnarray}}

\begin{document}

\title{The relativistic Aharonov-Bohm-Coulomb system with position-dependent mass}


\author{R. R. S Oliveira}
\email{rubensrso@fisica.ufc.br}
\affiliation{Universidade Federal do Cear\'a (UFC), Departamento de F\'isica,\\ Campus do Pici, Fortaleza - CE, C.P. 6030, 60455-760 - Brazil.}


\author{A. A. Araújo Filho}
\email{dilto@fisica.ufc.br}
\affiliation{Universidade Federal do Cear\'a (UFC), Departamento de F\'isica,\\ Campus do Pici, Fortaleza - CE, C.P. 6030, 60455-760 - Brazil.}


\author{R. V. Maluf}
\email{r.v.maluf@fisica.ufc.br}
\affiliation{Universidade Federal do Cear\'a (UFC), Departamento de F\'isica,\\ Campus do Pici, Fortaleza - CE, C.P. 6030, 60455-760 - Brazil.}


\author{C. A. S. Almeida}
\email{carlos@fisica.ufc.br}
\affiliation{Universidade Federal do Cear\'a (UFC), Departamento de F\'isica,\\ Campus do Pici, Fortaleza - CE, C.P. 6030, 60455-760 - Brazil.}

\date{\today}

\begin{abstract}

In this work, we study the Aharonov-Bohm-Coulomb (ABC) system for a relativistic Dirac particle with position-dependent mass (PDM). To solve our system, we use the $left$-$handed$ and $right$-$handed$ projection operators. Next, we explicitly obtain the eigenfunctions and the energy spectrum of the particle. We verify that these eigenfunctions are written in terms of the generalized Laguerre polynomials and the energy spectrum depends on the parameters $Z$, $\Phi_{AB}$ and $\kappa$. We notice that the parameter $\kappa$ has the function of increasing the values of the energy levels of the system. In addition, the usual ABC system is recovered when one considers the limit of constant mass ($\kappa\to{0}$). Moreover, also we note that even in the absence of ABC system ($Z=\Phi_{AB}=0$), the particle with PDM still has a discrete energy spectrum.

\end{abstract}

\maketitle

\section{Introduction}

The in literature, there a significant number of works that investigate the dynamics of particles with constant mass interacting with the vector potential of the Aharonov-Bohm (AB) effect \cite{Aharonov,Tonomura,Recher,Hagen1990} and the 2D Coulomb potential \cite{Neto,Kotov,Rathe,Dong}. A system described by the combination of these two potentials is so-called Aharonov-Bohm-Coulomb (ABC) system \cite{Nguyen,Draganascu,Doebner,Hagen}. In particular, the ABC system for spin-1/2 relativistic particles is studied in connection with the Feynman path integrals \cite{Lin,Bornales}, scattering \cite{Lin2000}, magnetic monopole \cite{Villalba1995,Hai}, spontaneous creation of fermions pairs \cite{Khalilov2009}, Coulomb impurity \cite{Nishida}, etc. Recently, the ABC system was applied in a graphene ring \cite{Jung} and studied together with the Dirac oscillator \cite{Oliveira}. 

Physical systems with effective mass, in special, with position-dependent mass (PDM) are particular interest in theoretical and experimental physics. For instance, using the Schr\"{o}dinger equation (SE) with PDM, we can investigate the electronic properties of semiconductors \cite{Krebs}, quantum well and quantum dots \cite{Harrison}, $^{3}$He clusters \cite{Barranco}, quantum liquids \cite{Saavedra}, etc. However, the relativistic extension this formalism it has the advantage of them to eliminate the problem of the ordering ambiguity between the mass and the momentum operator in the SE \cite{Plastino,Almeida}. In particular, using the Dirac equation (DE) with PDM, we can study problems involving scattering \cite{Alhaidari2007}, Coulomb field \cite{Alhaidari2004,Vakarchuk}, spin and pseudo-spin symmetry \cite{Ikhdair2010}, solid state physics \cite{Renan}, supersymmetry \cite{Ho2004}, PT-symmetry \cite{Jia,Castro,Mustafa}, infinite square well \cite{Alberto}, generalized uncertainty principle \cite{Pedram}, etc.

In this work, we investigate the relativistic quantum dynamics of an electrically charged Dirac particle with PDM in an ABC system in the $(2+1)$-dimensional Minkowski spacetime. To solve exact our problem, we use the $left$-$handed$ and $right$-$handed$ projection operators. Yet, we assume that the PDM be relevant for distances of the order of magnitude of the Compton wavelength $\lambda$, otherwise, we obtain the rest mass $m_0$ of the particle in the limit $\rho\to{\infty}$ or $\lambda\to{0}$, being $\rho$ the radial coordinate.

This paper is organized as follows. In Section \ref{sec2}, we introduce the DE in polar coordinates for an electrically charged particle with PDM in an ABC system. Next, we apply the $left$-$handed$ and $right$-$handed$ projection operators in the DE and we obtain a second order differential equation. In Section \ref{sec3}, we determine the eigenfunctions and the energy spectrum for the bound-states of the particle. In Section \ref{conclusion}, we present our conclusions.

\section{The Dirac equation with position-dependent mass in an Aharonov-Bohm-Coulomb system \label{sec2}}

The $(2+1)$-dimensional DE that governs the dynamics of an electrically charged particle with PDM in the presence of an external electromagnetic field $A_\mu$ reads as follows (Gaussian system in natural units $\hbar=c=1$) \cite{Greiner}
\ie [\gamma^{\mu}\Pi_{\mu}-m({\bf r})]\Psi(t,{\bf r})=0, \ \ (\mu=0,1,2),
\label{mass1}\fe
where $\gamma^{\mu}=(\gamma^0,\boldsymbol{\gamma})$ are the gamma matrices, $\Pi_{\mu}=p_\mu-qA_\mu$ is the kinetic momentum operator, being $p_{\mu}=i\partial_\mu$ the momentum operator, $q<0$ is the electric charge of the particle and $\Psi(t,{\bf r})$ is the two-component Dirac spinor. 
 
Now, we introduce the following  $left$-$handed$ and $right$-$handed$ projection operators \cite{Auvil}
\ie P_{L}=\frac{1}{2}(\mathbb{I}_{2\times 2}-\gamma^{5}), \ \ P_{R}=\frac{1}{2}(\mathbb{I}_{2\times 2}+\gamma^{5}),
\label{operators1}\fe
which satisfy the properties $P^{2}_{L}=P_{L}, P^{2}_{R}=P_{R},\{P_{L},P_{R}\}=0, P_{L}+P_{R}=\mathbb{I}_{2\times 2}$ and $P_{R}\gamma^{\mu}=\gamma^{\mu}P_{L}$, where $\gamma^5=\gamma_5=\sigma_1$ . Then, applying $P_L$ in Eq. \eqref{mass1} and defining the $left$-$handed$ and $right$-$handed$ spinors as $\Psi_L(t,{\bf r})=P_L\Psi(t,{\bf r})$ and $\Psi_R(t,{\bf r})=P_R\Psi(t,{\bf r})$, we get
\ie \Psi_L(t,{\bf r})=\frac{1}{m({\bf r})}\gamma^{\mu}\Pi_{\mu}\Psi_R(t,{\bf r}).
\label{mass2}\fe

The above relation allows us to write the original Dirac spinor in the form
\ie \Psi(t,{\bf r})=\frac{1}{m({\bf r})}[\gamma^{\mu}\Pi_{\mu}+m({\bf r})]\Psi_R(t,{\bf r}),
\label{mass3}\fe
where we used $\Psi(t,{\bf r})=\Psi_L(t,{\bf r})+\Psi_R(t,{\bf r})$.

Substituting the spinor \eqref{mass3} into Eq. \eqref{mass1}, we obtain
\ie [\gamma^{\mu}\Pi_{\mu}-m({\bf r})][\gamma^{\mu}\Pi_{\mu}+m({\bf r})]\Psi_R(t,{\bf r})=0.
\label{mass4}\fe

Adopting now the polar coordinates system $(t,\rho,\theta)$ where the metric tensor is given by $g^{\mu\nu}$=diag$(1,-1,-\rho^2)$ \cite{Villalba}, being $\rho=\sqrt{x^2+y^2}>0$ the radial coordinate and $0\le\theta\le2\pi$ the azimutal coordinate, Eq. \eqref{mass4} becomes
\ie A^{-}A^{+}\Psi_R(t,\rho,\theta)=0,
\label{mass5}\fe
where the operators $A^{\mp}$ are defined in the form
\ie A^{\mp}=\left[\gamma^0\left(i\frac{\partial}{\partial t}-qA_0\right)+i\gamma^{\rho}\frac{\partial}{\partial\rho}+\gamma^{\theta}\left(\frac{i}{\rho}\frac{\partial}{\partial\theta}+qA_{\theta}\right)\mp m(\rho)\right],
\label{operators2}\fe
with $\gamma^0=\beta$, $\gamma^{\rho}=\boldsymbol{\gamma}\cdot\hat{e}_{\rho}=\gamma^{1}\cos\theta+\gamma^{2}\sin\theta$ and $\gamma^{\theta}=\boldsymbol{\gamma}\cdot\hat{e}_{\theta}=-\gamma^{1}\sin\theta+\gamma^{2}\cos\theta$.

Here, we are explicitly assuming that the radial component of the vector potential is null ($A_{\rho}=0$). Also, through a similarity transformation given by unitary operator $U(\theta)=e^{-\frac{i\theta\sigma_3}{2}}$, we can reduce the matrices $\gamma^{\rho}$ and $\gamma^{\theta}$ to the matrices $\gamma^{1}$ and $\gamma^{2}$ as follows \cite{Villalba}
\ie U^{-1}(\theta)\gamma^{\rho}U(\theta)=\gamma^{1}, \  U^{-1}(\theta)\gamma^{\theta}U(\theta)=\gamma^{2}.
\label{mass6}\fe

Since we are working in a $(2+1)$-dimensional Minkowski spacetime, it is convenient define the Dirac matrices $\boldsymbol{\gamma}=(\gamma^1,\gamma^2)=(-\gamma_1,-\gamma_2)$ and $\gamma^0$ in terms of the Pauli matrices, i.e., $\gamma_1=\sigma_3\sigma_1$, $\gamma_2=\sigma_3\sigma_2$ and $\gamma^0=\sigma_3$ \citep{Greiner,Villalba}. Therefore, using this information and the relations \eqref{mass6}, we rewrite Eq. \eqref{mass5} in the form
\ie B^{-}B^{+}\psi_R(t,\rho,\theta)=0,
\label{mass7}\fe
where
\ie B^{\mp}=\left[\sigma_3\left(i\frac{\partial}{\partial t}-qA_0\right)+\sigma_2\frac{\partial}{\partial\rho}+i\sigma_1\left(\frac{i}{\rho}\frac{\partial}{\partial\theta}+qA_{\theta}+\frac{\sigma_3}{2\rho}\right)\mp m(\rho)\right],
\label{operators3}\fe
\ie \psi_R(t,\rho,\theta)=U^{-1}(\theta)\Psi_R(t,\rho,\theta)
\label{spinor1}.\fe

Let us now consider the configurations of the vector potential of the AB effect and of the 2D Coulomb potential. Explicitly, these configurations are given in the form \cite{Nguyen,Draganascu,Doebner,Hagen,Jung}
\ie {\bf A}=A_\theta\hat{e}_{\theta}=\frac{\Phi}{2\pi\rho}\hat{e}_{\theta}, \ (\rho>a),
\label{vectorpotential}\fe
\ie V=qA_0=-\frac{Ze^2}{\rho}, \ (q=-e),
\label{Coulomb}\fe
where $\Phi=\pi a^2 B=const$ is the magnetic flux in the region intern of a solenoid of radius $a$ electrically charged with a total charge $Ze>0$, being $Z$ the atomic number. 

With respect to the configuration of the variable mass, we consider the following PDM
\ie m(\rho)=m_0+\frac{\kappa}{\rho}
\label{mass},\fe
where $m_0$ is the rest mass of the particle and $\kappa>0$ is a real parameter. In special, the PDM \eqref{mass} is the two-dimensional version of a spherically symmetrical singular mass distribution worked in Refs. \cite{Alhaidari2004,Ikhdair2010,Vakarchuk}. In agreement with Ref. \cite{Alhaidari2004}, the parameter $\kappa$ can be defined in the form $\kappa=m_0\mu\lambda^{2}$, where $\lambda$ is the Compton wavelength of the particle and $\mu$ is a real scale parameter with length inverse dimension. However, as the PDM \eqref{mass} diverges at the origin, the parameter $\nu$ can be interpreted as a renormalization scale in quantum field theory (QFT) to eliminate the ultraviolet divergences that appear in high energy physics \cite{Alhaidari2004}.

Therefore, using the configurations \eqref{vectorpotential}, \eqref{Coulomb} and \eqref{mass}, we transform Eq. \eqref{mass7} as follows
\ie \left[\frac{\partial^{2}}{\partial\rho^{2}}+\frac{1}{\rho}\frac{\partial}{\partial\rho}-\frac{1}{4\rho^2}-\frac{{\bf\Gamma}({\bf\Gamma}-1)}{\rho^{2}}+\frac{{\bf\Delta}}{\rho}+{\bf\Sigma}\right]\psi_R(t,\rho,\theta)=0
\label{mass8}, \fe
where we define the following operators
\ie {\bf\Gamma}^{2}\equiv{\left[\left(L_z+\frac{e\Phi}{2\pi}\right)^{2}-Z^2 e^4+\kappa^2\right]}, \ \ {\bf\Gamma}\equiv{\left(iL_z\sigma_2\sigma_1+Ze^2\sigma_3\sigma_2-\kappa\sigma_2+ie\Phi_{AB}\sigma_2\sigma_1\right)}
\label{operators4}, \fe
\ie {\bf\Delta}\equiv{\left(2iZe^2\frac{\partial}{\partial t}-2m_0\kappa\right)}, \ \ {\bf\Sigma}\equiv{\left(-\frac{\partial^2}{\partial t^2}-m^{2}_{0}\right)},\ \ L_z=-i\frac{\partial}{\partial\theta}
\label{operators5}. \fe

Writing the two-component Dirac spinor in the form \cite{Villalba}
\ie \psi_{R}(t,\rho,\theta)=\frac{e^{i(m_{l}\theta-Et)}}{\sqrt{2\pi}}\left(
           \begin{array}{c}
            \phi_+(\rho) \\
             \phi_-(\rho) \\
           \end{array}
         \right),  \ (m_{l}=\pm1/2,\pm3/2,\ldots)
\label{spinor2},\fe
Eq. \eqref{mass8} becomes compacted in the following differential equation
\ie \left[\frac{d^{2}}{d\rho^{2}}+\frac{1}{\rho}\frac{d}{d\rho}-\frac{\gamma^2_s}{\rho^{2}}+\frac{(2Z\alpha E-2m_0\kappa)}{\rho}+E^2-m_{0}^2\right]\phi_s(\rho)=0, \ \ (s=\pm 1),
\label{mass9} \fe
where
\ie \gamma_s\equiv\sqrt{(m_{l}+\Phi_{AB})^2-Z^2\alpha^2+\kappa^2}-\frac{s}{2}
\label{mass10}, \fe
being $\phi^{s}(\rho)$ real radial functions, $E$ is the relativistic total energy of the particle, $m_{l}$ is the orbital magnetic quantum number, $\Phi_{AB}=\frac{\Phi}{\Phi_0}>0$ is the geometric phase AB, being $\Phi_0=\frac{2\pi}{e}$ the magnetic flux quantum and $\alpha=e^2\cong\frac{1}{137}$ is the Sommerfeld fine structure constant. 

\section{Bound-state solutions and energy spectrum \label{sec3}} 

In order to solve Eq. \eqref{mass10}, we will introduce now a new dimensionless variable given by $z=2\eta\rho$, where $\eta=\sqrt{m^2_0-E^2}$, being $m^2_0>E^2$. Thereby, Eq. \eqref{mass10} becomes
\ie \left[\frac{d^{2}}{dz^{2}}+\frac{1}{z}\frac{d}{dz}-\frac{\gamma^2_s}{z^2}+\frac{z_0}{z}-\frac{1}{4}\right]\phi_s(z)=0
\label{solution1}, \fe
where
\ie z_0\equiv\frac{(Z\alpha E-m_0\kappa)}{\eta}
\label{solution2}.\fe

Analyzing the asymptotic behavior of Eq. \eqref{solution1} for $z\to{0}$ and $z\to{\infty}$, we obtain
\ie \phi_s(z)=C_s z^{\vert \gamma_s\vert}e^{-\frac{z}{2}}R_s(z)
\label{solution3},\fe 
where $C_s$ are normalization constants and $R_s(z)$ are unknown functions to be determined.

In this way, substituting \eqref{solution3} into Eq. \eqref{solution1}, we have
\ie z\frac{d^{2}R^{s}(z)}{dz^{2}}+\left(2\vert\gamma_s\vert+1-z\right)\frac{dR^{s}(z)}{dz}-\left(\frac{2\vert\gamma_s\vert+1}{2}-z_0\right)R_s(z)=0
\label{solution4}.\fe

It is not difficult to note that Eq. \eqref{solution4} has the form of a generalized Laguerre equation, whose solution are the generalized Laguerre polynomials $R_s(z)=L^{2\vert\gamma_s\vert}_n(z)$ \cite{Abramowitz}. Besides that, to $\phi^s(z)$ be a normalizable solution, we must impose that the parameter $\vert\gamma_s\vert+\frac{1}{2}-z_0$ to be equal to a non-positive integer number $-n$ ($n=0,1,2,\ldots$). Therefore, using this condition and the relation \eqref{solution2}, we obtain the following energy spectrum for the Dirac particle with PDM in an ABC system
\ie E^+_{n,m_l}=\frac{Z\alpha m_0\kappa}{[(n_s+\gamma)^2+(Z\alpha)^2]}+m_0\sqrt{\frac{(Z\alpha\kappa)^2}{[(n_s+\gamma)^2+(Z\alpha)^2]^2}+\frac{(n_s+\gamma)^2-\kappa^2}{[(n_s+\gamma)^2+(Z\alpha)^2]}},
\label{energyspectrum}\fe
where $n_s=n+\frac{1-s}{2}$ is a quantum number and $\gamma\equiv\sqrt{(m_{l}+\Phi_{AB})^2-Z^2\alpha^2+\kappa^2}>0$. We see that the energy spectrum \eqref{energyspectrum} explicitly depends on the parameters $Z$ and $\Phi_{AB}$ that characterize the ABC system and of the parameter $\kappa$ that characterizes the PDM. Due to the presence of the term $s=\pm 1$, we see that the upper component of the Dirac spinor has energy eigenvalues slightly larger than the lower component. In addition, the negative signal in \eqref{energyspectrum} was excluded because for an positively charged solenoid ($Z\alpha>0$), the negative energy states $E<0$ does not satisfy the relation: $z_0=n+\vert\gamma_s\vert+\frac{1}{2}>0$.

Comparing the energy spectrum \eqref{energyspectrum} with the literature, we see that in the limit of the constant mass ($\kappa\to{0}$), or, nonrenormalization limit \cite{Alhaidari2004}, the energy spectrum of the usual ABC system is recovered \cite{Bornales}. Moreover, whether in the conditions $\kappa\neq{0}$ or $\kappa=0$, the energy spectrum of the ABC system still it has degeneracy finite. Also interesting to note that even in the absence of the ABC system ($Z=\Phi_{AB}=0$), the Dirac particle with PDM still has a discrete energy spectrum. In this sense, we can interpret the term of the PDM that varies spatially as a type of scalar coupling in the DE, where we have the Lorentz-scalar potential $V_s(\rho)=\frac{\kappa}{\rho}$, and whose energy spectrum is given as follows: $E_{n,m_l}=\pm m_0\sqrt{1-\kappa^2(n_s+\sqrt{m^2_l+\kappa^2})^{-2}}$.

Now, let us concentrate on the form of the original Dirac spinor. Substituting the variable $z=2\eta\rho$ in the radial function \eqref{solution3}, the spinor \eqref{spinor2} becomes
\ie \psi_{R}(t,\rho,\theta)=e^{i(m_{l}\theta-Et)}\left(
           \begin{array}{c}
            \bar{C}_+ \rho^{\vert\gamma_+\vert}e^{-\eta\rho}L^{2\vert\gamma_+\vert}_n(2\eta\rho) \\
             \bar{C}_- \rho^{\vert\gamma_-\vert}e^{-\eta\rho}L^{2\vert\gamma_-\vert}_n(2\eta\rho) \\
           \end{array}
         \right)
\label{spinor3},\fe
where
\ie \bar{C}_s\equiv\frac{C_s(2\eta)^{\vert\gamma_s\vert}}{\sqrt{2\pi}}
\label{constant}. \fe

Now, substituting the spinor \eqref{spinor1} in the spinor \eqref{mass3} and using the relations \eqref{mass7}, we obtain
\ie \Psi=\frac{1}{m(\rho)}U\left[\sigma_3\left(i\frac{\partial}{\partial t}+\frac{Ze^2}{\rho}\right)+m(\rho)+\sigma_2\left(\frac{\partial}{\partial\rho}+\frac{1}{2\rho}\right)+i\sigma_1\left(\frac{i}{\rho}\frac{\partial}{\partial\theta}-\Phi_{AB}\right)\right]\psi_R
\label{spinor4}. \fe

Therefore, substituting the spinor \eqref{spinor3} in \eqref{spinor4}, the original Dirac spinor is written in the form
\ie \Psi_{n,m_l}(t,\rho,\theta)=\left(
           \begin{array}{c}
            e^{i[(m_l-\frac{1}{2})\theta-Et]}[F_+(\rho)+iG_-(\rho)] \\
             e^{i[(m_l+\frac{1}{2})\theta-Et]}[F_-(\rho)-iG_+(\rho)] \\
           \end{array}
         \right)
\label{spinor5},\fe
where
\ie F_s(\rho)=\frac{\bar{C}_s}{m(\rho)}e^{-\eta\rho}\rho^{\vert\gamma_s\vert}L^{2\vert\gamma_s\vert}_n\left(2\eta\rho\right)\left(E+m_0+\frac{(\kappa-sZe^2)}{\rho}\right)
\label{solution5},\fe
\ie G_s(\rho)=\frac{\bar{C}_s}{m(\rho)}e^{-\eta\rho}\rho^{\vert\gamma_s\vert}\left[\ L^{2\vert\gamma_s\vert}_n\left(2\eta\rho\right)\left(\eta+\frac{(sm_l-\frac{1}{2}-\vert\gamma_s\vert+s\Phi_{AB})}{\rho}\right)+L^{2\vert\gamma_s\vert+1}_{n-1}\left(2\eta\rho\right)\right]\
\label{solution6}.\fe

\section{Conclusion\label{conclusion}}

In this paper, we study the $(2+1)$-dimensional DE for an electrically charged particle with PDM in an ABC system. Next, we applied the  $left$-$handed$ and $right$-$handed$ projection operators in the DE and we obtain a second order differential equation. After analyzing the asymptotic behavior this differential equation for $z\to{0}$ and $z\to{\infty}$, we obtain a generalized Laguerre equation. As results, we observed that the energy spectrum of the particle explicitly depends on the parameters $Z$ and $\Phi_{AB}$ that characterize the ABC system and of the parameter $\kappa$ that characterizes the PDM. We verify that the parameter $\kappa$ has the finality of increasing the values of the energy levels of the system. We observed also that in the limit of the constant mass ($\kappa\to{0}$), the energy spectrum of the usual ABC system is recovered. Moreover, also we note that even in the absence of the ABC system ($Z=\Phi_{AB}=0$), the Dirac particle with PDM still has a discrete energy spectrum. In view of this, we can interpret the PDM as a type of scalar coupling in the DE.

\section*{Acknowledgments}
\hspace{0.5cm}The authors would like to thank the Funda\c{c}\~{a}o Cearense de apoio ao Desenvolvimento Cient\'{\i}fico e Tecnol\'{o}gico (FUNCAP) the Coordena\c{c}\~ao de Aperfei\c{c}oamento de Pessoal de N\'ivel Superior (CAPES), and the Conselho Nacional de Desenvolvimento Cient\'{\i}fico e Tecnol\'{o}gico (CNPq) for financial support.

\end{document}